\documentstyle[amssymb,preprint,aps]{revtex}
\tightenlines

\begin{document}
\draft
\title{
COMPACT AND LOOSELY BOUND STRUCTURES IN LIGHT NUCLEI}
\author{O.L.Savchenko, A.I.Steshenko}
\address{Bogolyubov Institute for Theoretical Physics of the NAS of Ukraine \\
Metrolohichna Str., 14b, Kyiv-143, 03143 Ukraine}
\date{\today }
\maketitle

\begin{abstract}
 A role of different components in the wave function of
the weakly bound light nuclei states was studied within the framework of
the cluster model, taking into account of orbitals
"polarization". It was shown that a limited number of structures
associated with the different modes of nucleon motion can be of
great importance for such systems. Examples of simple and quite
flexible trial wave functions are given for the nuclei $^8$Be,
$^6$He. Expressions for the microscopic wave functions of
these nuclei were found and used for the calculation of basic
nuclear characteristics, using well known central-exchange
nucleon-nucleon potentials.
\end{abstract}
\pacs{64.60.Cn, 81.30.Bx, 75.10.H}

\section{Introduction.}

 Theoretical consideration of strongly interacting
many-particle systems, such as the atomic nuclei, has a number of
difficulties, one of which is constructing of trial wave
function. The wave function used should be characterized by proper
set of the definite quantum numbers and corresponding symmetries,
taking into account the most important physical modes of
nucleon motion. It is usually assumed that structure of the
constructed wave function is well known in the limitsd cases,
i.e. in the case where a distance between fragments is very large
or when we have a compact one-centre system.
Hence, the main attention is paid to determination of correct
interpolation expression for the wave function in the intermediate
region.

Many authors have worked at simultaneous description of both
"compact" and "friable" structures (see, for example,
Ref.\cite{1,2}). Several decades ago due to activity of
Yu.A.Simonov, great attention was paid to the hyperspherical
function \cite{3,4,5} method (the K-harmonic method) which was
considered as a general solution method for the nuclear many-particle
problem. But it was shown that its practical
application is limited effectively to compact systems;
moreover, the results obtained \cite{6,7} hardly differ from the
results calculated using the simplest functions of the oscillator
shell model with one variation parameter.

Recently quite effective stochastic variational methods
\cite{8,9,10,11,12} for semirealistic nucleon-nucleon (NN)
potentials using a number of multicluster model basic functions
have been developed for the case of the lightest ($A\le 10$)
nuclei. They allow to find almost exact numerical solutions
of many particle problems for some variants of
NN-forces which are most frequently used in calculations of the
light nuclei structure. Unfortunately, only numerical estimations
of energies and radii are usually obtained due to complication of
the calculations. Thus, results of such microscopic computations
rarely can be used for estimation of other nuclear
characteristics. Therefore, it is desirable to have many-particle
wave function in evident form for real efficiency of the method.
Moreover, if we have the wave function in the explicit form, we
will be able to use it in calculations of different nuclear
processes within the nuclear system.

To construct simple and quite flexible
microscopic trial wave function of the nucleus which correctly
describes both compact and weakly bound states, it is of prime
importance at studing of exotic nuclei with anomalous ratio N/Z.

It is known that deformation effects are significant in nuclei
with $A>4$. But concrete calculations are usually performed on the
basis of spherical symmetry for cluster wave functions and
detailed function for cluster-cluster motion. Recently, we have
tried \cite{13} to take into consideration the deformation effects
in the framework of the many-particle oscillator shell model by
using the variation principle (without taking into account the
cluster modes). In the present paper we describe microscopic
model which employs orbital "polarization" as well as cluster
degrees of the nucleon motion and exact angular projection. The
model can be used for description of both compact and weakly bound
states in the light and medium light atomic nuclei.

Below we briefly outline the construction of the trial wave
function (section \ref{sect2}) and give all matrix elements (m.e.)
of the basic physical operators needed for our computation
(section \ref{sect3}). The variational calculation results for the
nuclei $^8$Be, $^6$He are given in section \ref{sect4}. \\[1mm]

\section{Many-particle Hamiltonian and trial wave functions.} \label{sect2}

Let us split considered A-nucleon system in two subsystems
($A=A_1+A_2$) in such way that nucleon-nucleon interaction inside
of each subsystem is a priori stronger then that between the two
nucleons in the different fragments. We introduce the two
fragments in order to simplify the account. In case if there are
more fragments, the generalization can be written in a similar
way. Note that nucleon motion mode defined by interaction between
subsystems will be the "softest". Analysis of the experimental
data for the considered nucleus allows to choose correct splitting
of A-nucleon system into fragments. As a rule, the most
interesting "physics" at low energies is caused by the soft mode
which should be the most detailed in the framework of the chosen
model. For example, description of the A-nucleus weak-bound state
will be sufficient if we were able to obtain the basic nuclear
characteristics as well as the decay $A=A_1+A_2$ threshold on the
assumption that separated fragments $A_1$ and $A_2$ are known
compact nuclei.

In general, every microscopic model is associated with its own
effective NN-potential. The most known effective NN-potentials
have been constructed in the framework of the models based on the
shell model wave function with one variation parameter. It is
clear that using such NN-potentials in the more advanced models
can be uneffective. However, this usage can be quite correct if
NN-potential was built on the basis of the compact nuclei
properties (e.g., ones of the magic nuclei $^4$He, $^{16}$O or
$^{40}$Ca). The results of calculations \cite{13} for the magic
nuclei $^4$He, $^{16}$O are rather similar for both simple and
advanced model, therefore is valid to use mentioned above
NN-potentials for the "polarized" orbitals model. Let us further
consider the standard microscopic nuclear model \cite{14}  based
on the many-nucleon Hamiltonian with two-particle central exchange
interaction
\begin{equation}
\label{1} \hat V_{ij}(r)=\sum_{S,T=0,1} V_{2S+1,2T+1}(r)\cdot \hat
P_{ST}(ij) \ , \qquad r\equiv |\vec r_i-\vec r_j| \ ,
\end{equation}
where the radius dependence of the potential components can be
represented by Gaussian superposition, i.e.
\begin{equation}
\label{2} V_{2S+1,2T+1}(r)=\sum_{\nu =1}^{\nu_{pot}}
V_{2S+1,2T+1}^{[\nu ]} \cdot
\exp\left(-\frac{r^2}{\mu^2_{\nu}}\right) \ ;
\end{equation}
projection operator $\hat P$ in eq.(\ref{1}) cut out the state of
nucleon pair ($ij$) with defined spin $S$ and isospin $T$.

When fragments are separated enough, the Hamiltonian $\hat H_A$
should turn into the sum of separated subsystem Hamiltonians plus
Coulomb repulsion between them due to the short-ranged nature of
the nuclear forces
\begin{equation}
\label{3} \hat H_A \Rightarrow \hat H_{A_1}+\hat H_{A_2}+
 \frac{Z_1Z_2e^2}{|\vec R_1-\vec R_2|} \ , \qquad
 \vec R_l=\frac{1}{A_l}\sum_{i=1}^{A_l} \vec r_i \ ; \qquad
 l=1,2,
\end{equation}
i.e., the limiting cases are known.

The behaviour of the system in the intermediate range where
altitude of vector $\vec S=\vec R_1-\vec R_2$ significantly
differs from zero, but is not large enough to neglect the nuclear
interaction between particles in the different clusters, is open
question. Hence, it is necessary to define dependency of the
center-of-mass kinetic energy operator $\hat T_{c.m.}$ on the
stage of separation of the A-nucleus into fragments. We simulate
$\hat T_{c.m.}$ in this range by expression

\begin{equation}
\label{4} \hat T_{c.m.}(\alpha_S)=-\frac{\hbar^2}{2Am}
  \Bigl[1+\alpha_S\left(\frac{A}{A_1}-1\right)\Bigr]\cdot
  \left(\sum_{i=1}^{A_1}\vec{\nabla}_i\right)^2-
\end{equation}

$$
  -\frac{\hbar^2}{2Am}
  \Bigl[1+\alpha_S\left(\frac{A}{A_2}-1\right)\Bigr]\cdot
  \left(\sum_{j=A_1+1}^{A}\vec{\nabla}_j\right)^2-
  \frac{\hbar^2}{Am}\left(1-\alpha_S\right)\cdot
  \sum_{i=1}^{A_1}\sum_{j=A_1+1}^{A} \vec{\nabla}_i\vec{\nabla}_j \ .
$$
In our calculations the value of the parameter $\alpha_S$
$([0\le\alpha_S\le 1])$ in eq.(\ref{4}) was defined by the degree
of importance of the Pauli principle in the A-nucleon system.
Thus, $\alpha_S=0$ if antisymmetrization acts on the variables of
all A nucleons and $\alpha_S=1$ in the other limiting case when
the Pauli principle acts only on the nucleons inside the clusters.

A total trial wave function $\Psi_A$ is represented by
superposition of a small number of determinant functions
\begin{equation}
\label{5} \Psi_A(1,2,\dots ,A)=\sum_{\nu}w_{\nu}\cdot\Psi_{\nu} \
,
\end{equation}
where $\Psi_{\nu}$-components (normalized, but generally not
orthogonal) are well known Slater determinants
\begin{equation}
\label{6} \Psi_{\nu}=\frac{f_{\nu}}{\sqrt{A!}}
   \det\left[\phi_i^{[\nu ]}(j)\right] \ ,\qquad
   f_{\nu}=\frac{1}{\sqrt{\det\left[
   <\phi_i^{[\nu ]}|\phi_j^{[\nu ]}>\right]}} \ ,
\end{equation}
filled by one-particle orbitals which have unificated form:
$$ \phi_i(j)=\psi_{\vec n_i}(\vec r_j;\vec a_i,\vec R_i)\cdot
  \chi_{\sigma\tau}(j) \ ; \qquad
  \sum_{\sigma\tau}\int |\phi_i|^2 d\vec r=1 \ ,
$$

\begin{equation}
\label{7} \psi_{\vec n}(\vec r;\vec a,\vec R)=\psi_{n^x}(x;a,R^x)
   \psi_{n^y}(y;b,R^y)\psi_{n^z}(z;c,R^z) \ ,
\end{equation}

$$ \psi_n(x;a,R)=\frac{1}{\sqrt{a\sqrt{\pi}}\cdot N(n;R/a)}\cdot
   \left(\frac{x}{a}\right)^n\cdot\exp\left\{
   -\frac{(x-R)^2}{2a^2}\right\} \ ,
$$

$$ N(n;R/a)=\sqrt{\left(\frac{R}{a}\right)^{2n}\cdot
  F\left( -n,-n+\frac{1}{2};0;\frac{a^2}{R^2}\right)} \ ,
$$
where $\chi_{\sigma\tau}$ is the spin-isospin function and
$F(\alpha ,\beta ;\gamma ;z)$ is the Gauss hyperspherical function
\cite{15}.

Every component $\Psi_{\nu}$ in eq.(\ref{5}) represents some
characteristic structure or configuration of the considered system
and is associated to some type of nuclear motion. We hope that the
wave function $\Psi_A$ constructed in such way will be flexible
enough at the reasonable number of the variation parameters
$\{\vec a_i,\vec R_i\}$.\\[1mm]

\section{Overlap integrals.} \label{sect3}

In order to perform the variational calculations it is necessary
to obtain the matrix elements of some physical operators. Needed
overlap integrals can be expressed in terms of "partial" m.e. on
the basis of the function set (\ref{7}) using known determinant
function technique \cite{16}.

At the beginning, we consider overlap integral of the functions
$\Psi_{\nu}$ with unity, i.e.
\begin{equation}
\label{8} <\Psi_{\nu}|\Psi_{\mu}>=f_{\nu}f_{\mu}\cdot
   \det\left[ <\phi_i^{[\nu ]}|\phi_j^{[\mu ]}>\right]\equiv
   f_{\nu}f_{\mu}\cdot\det\left[ (B_{[\nu ,\mu ]})_{ij}\right] \ ,
\end{equation}
where the matrix $B_{[\nu ,\mu ]}$ is defined by
\begin{equation}
\label{9} (B_{[\nu ,\mu
]})_{ij}=<\chi_{\sigma_i\tau_i}|\chi_{\sigma_j\tau_j}>
   \int\psi_{\vec n_i}^{[\nu ]}(\vec r;\vec a_i,\vec R_i)
   \psi_{\vec n_j}^{[\mu ]}(\vec r;\vec a_j,\vec R_j) d\vec r=
\end{equation}
$$ =<\chi_{\sigma_i\tau_i}|\chi_{\sigma_j\tau_j}>\cdot
   g_{n_i^xn_j^x}(a_iR_i^x;a_jR_j^x)
   g_{n_i^yn_j^y}(b_iR_i^y;b_jR_j^y)
   g_{n_i^zn_j^z}(c_iR_i^z;c_jR_j^z) \ .
$$

Evident expression of the partial m.e.
\begin{equation}
\label{10}
g_{n_1n_2}(a_1R_1;a_2R_2)\equiv\int\limits_{-\infty}^{\infty}
   \psi_{n_1}(x;a_1,R_1)\psi_{n_2}(x;a_2,R_2)dx
\end{equation}
is given in the Appendix (see eq.(\ref{A1})).

The general expression \cite{16} for one-particle operator matrix
elements
\begin{equation}
\label{11} \frac{1}{<\Psi_{\nu}|\Psi_{\mu}>}\cdot
   <\Psi_{\nu}|\sum_{i=1}^A\hat \Omega_i|\Psi_{\mu}>=
   \sum_{i,j=1}^A <\phi_i^{[\nu ]}|\hat \Omega_1|\phi_j^{[\mu ]}>
   \cdot\left( B^{-1}_{[\nu ,\mu ]}\right)_{ij}
\end{equation}
allows to obtain: \\ 1) m.e. for nuclear matter distribution $\rho
(\vec r)$
\begin{equation}
\label{12} \frac{1}{<\Psi_{\nu}|\Psi_{\mu}>}\cdot
   <\Psi_{\nu}|\sum_{i=1}^A\delta(\vec r-\vec r_i)|\Psi_{\mu}>=
   \sum_{i,j=1}^A
   \psi_{\vec n_i}^{[\nu ]}(\vec r;\vec a_i,\vec R_i)
   \psi_{\vec n_j}^{[\mu ]}(\vec r;\vec a_j,\vec R_j)\cdot
   \left( B^{-1}_{[\nu ,\mu ]}\right)_{ij} \ ;
\end{equation}
2) m.e. for kinetic energy operator
\begin{equation}
\label{13} \frac{1}{<\Psi_{\nu}|\Psi_{\mu}>}\cdot
   <\Psi_{\nu}|-\sum_{i=1}^A\frac{\hbar^2}{2m}\vec{\nabla}^2_i|\Psi_{\mu}>=
   \sum_{i,j=1}^A <\psi_{\vec n_i}^{[\nu ]}
   |\hat t_x+\hat t_y+\hat t_z|\psi_{\vec n_j}^{[\mu ]}>
   \cdot\left( B^{-1}_{[\nu ,\mu ]}\right)_{ij} \ ,
\end{equation}
where
\begin{equation}
\label{14} <\psi_{\vec n_i}|\hat t_x|\psi_{\vec n_j}>\equiv
  <\psi_{\vec n_i}^{[\nu ]}(\vec r;\vec a_i,\vec R_i)
   |-\frac{\hbar^2}{2m}\frac{\partial^2}{\partial x^2}|
   \psi_{\vec n_j}^{[\mu ]}(\vec r;\vec a_j,\vec R_j)>=
\end{equation}

$$
=g_{n_i^yn_j^y}(b_iR_i^y;b_jR_j^y)g_{n_i^zn_j^z}(c_iR_i^z;c_jR_j^z)
   \cdot t_{n_i^xn_j^x}(a_iR_i^x;a_jR_j^x) \ .
$$
The analogous expressions can be written for $<\psi_{\vec
n_i}|\hat t_y|\psi_{\vec n_j}>$ (substitution $x\rightarrow y$;
$a\rightarrow b$) and for $<\psi_{\vec n_i}|\hat t_z|\psi_{\vec
n_j}>$ (substitution $x\rightarrow z$; $a\rightarrow c$). The
matrix elements $t_{n'n}(a',R';a,R)$ employed in eq.(\ref{14}) are
given in the Appendix (see eq.(\ref{A2})). In
eqs.(\ref{11})--(\ref{13}) and below the inverse matrix
$$ B^{-1}_{[\nu ,\mu ]}=
   \Biggl{\|}\left( B^{-1}_{[\nu ,\mu ]}\right)_{ij}\Biggr{\|} \ .
$$
is used.

Further, let us use the general formula \cite{16} for two-particle
operators $\sum_{i,j=1}^A\hat{\Omega}_{ij}$
$$ \frac{1}{<\Psi_{\nu}|\Psi_{\mu}>}\cdot
   <\Psi_{\nu}|\sum_{i,j=1}^A\hat \Omega_{ij}|\Psi_{\mu}>=
   \sum_{k',l',k,l=1}^A <\phi_{k'}^{[\nu ]}\phi_{l'}^{[\nu ]}
   |\hat \Omega_{12}|\phi_k^{[\mu ]}\phi_l^{[\mu ]}>\times
$$

\begin{equation}
\label{15}
   \times\Biggl\{\left( B^{-1}_{[\nu ,\mu ]}\right)_{kk'}
   \left( B^{-1}_{[\nu ,\mu ]}\right)_{ll'}-
   \left( B^{-1}_{[\nu ,\mu ]}\right)_{kl'}
   \left( B^{-1}_{[\nu ,\mu ]}\right)_{lk'}\Biggr\} \ .
\end{equation}
In the case of the central exchange NN-potential (\ref{1}),
(\ref{2}) we can write for even-even nuclei after summing up on
the spin-isospin variables
$$ \frac{1}{<\Psi_{\nu}|\Psi_{\mu}>}\cdot
   <\Psi_{\nu}|\sum_{i,j=1}^A\hat V_{ij}|\Psi_{\mu}>=
$$

$$ =\sum_{i',i=1}^Z\sum_{j',j=1}^Z\sum_{l=1}^{l_{pot}}
   <\psi_{\vec n_{i'}}^{[\nu ]}(1)\psi_{\vec n_{j'}}^{[\nu ]}(2)
   |\exp\left\{-\frac{|\vec r_1-\vec r_2|^2}{\mu_l^2}\right\}|
   \psi_{\vec n_i}^{[\mu ]}(1)\psi_{\vec n_j}^{[\mu ]}(2)>\times
$$

$$
   \times\Biggl\{X_d(l)\cdot
   \left( B^{-1}_{[\nu ,\mu ]}\right)_{ii'}
   \left( B^{-1}_{[\nu ,\mu ]}\right)_{jj'}-X_{ex}(l)\cdot
   \left( B^{-1}_{[\nu ,\mu ]}\right)_{ij'}
   \left( B^{-1}_{[\nu ,\mu ]}\right)_{ji'}\Biggr\} +
$$

\begin{equation}
\label{16} +\sum_{i',i=1}^Z\sum_{j',j=1}^N\sum_{l=1}^{l_{pot}}
   <\psi_{\vec n_{i'}}^{[\nu ]}(1)\psi_{\vec n_{j'}}^{[\nu ]}(2)
   |\exp\left\{-\frac{|\vec r_1-\vec r_2|^2}{\mu_l^2}\right\}|
   \psi_{\vec n_i}^{[\mu ]}(1)\psi_{\vec n_j}^{[\mu ]}(2)>\times
\end{equation}

$$
   \times\Biggl\{Y_d(l)\cdot
   \left( B^{-1}_{[\nu ,\mu ]}\right)_{ii'}
   \left( B^{-1}_{[\nu ,\mu ]}\right)_{jj'}-Y_{ex}(l)\cdot
   \left( B^{-1}_{[\nu ,\mu ]}\right)_{ij'}
   \left( B^{-1}_{[\nu ,\mu ]}\right)_{ji'}\Biggr\} +
$$

$$ +\sum_{i',i=1}^N\sum_{j',j=1}^N\sum_{l=1}^{l_{pot}}
   <\psi_{\vec n_{i'}}^{[\nu ]}(1)\psi_{\vec n_{j'}}^{[\nu ]}(2)
   |\exp\left\{-\frac{|\vec r_1-\vec r_2|^2}{\mu_l^2}\right\}|
   \psi_{\vec n_i}^{[\mu ]}(1)\psi_{\vec n_j}^{[\mu ]}(2)>\times
$$

$$
   \times\Biggl\{X_d(l)\cdot
   \left( B^{-1}_{[\nu ,\mu ]}\right)_{ii'}
   \left( B^{-1}_{[\nu ,\mu ]}\right)_{jj'}-X_{ex}(l)\cdot
   \left( B^{-1}_{[\nu ,\mu ]}\right)_{ij'}
   \left( B^{-1}_{[\nu ,\mu ]}\right)_{ji'}\Biggr\} \ ,
$$
where
\begin{equation}
\label{17} X_d(l)=\frac{3V_{33}^{(l)}+V_{13}^{(l)}}{2} \ ,\qquad
X_{ex}(l)=\frac{3V_{33}^{(l)}-V_{13}^{(l)}}{2} \ ,
\end{equation}

$$ Y_d(l)=\frac{3V_{33}^{(l)}+3V_{31}^{(l)}+V_{13}^{(l)}+
   V_{11}^{(l)}}{2} \ ,\qquad
Y_{ex}(l)=\frac{3V_{33}^{(l)}-3V_{31}^{(l)}-V_{13}^{(l)}+
   V_{11}^{(l)}}{2} \ .
$$

Due to the Gaussian dependence of the considered potential the
partial matrix elements between the basis functions (\ref{7}) are
factorized on three terms, i.e.
$$ <\psi_{\vec n_{i'}}(1)\psi_{\vec n_{j'}}(2)
   |\exp\left\{-\frac{|\vec r_1-\vec r_2|^2}{\mu^2}\right\}|
   \psi_{\vec n_i}(1)\psi_{\vec n_j}(2)>=
$$

\begin{equation}
\label{18}
=j_{12}(n_{i'}^x a_{i'} R_{i'}^x, n_{j'}^x a_{j'} R_{j'}^x;\mu ;
        n_{i}^x a_{i} R_{i}^x, n_{j}^x a_{j} R_{j}^x)\times
\end{equation}
$$ \times j_{12}(n_{i'}^y b_{i'} R_{i'}^y, n_{j'}^y b_{j'}
R_{j'}^y;\mu ;
          n_{i}^y b_{i} R_{i}^y, n_{j}^y b_{j} R_{j}^y)
       j_{12}(n_{i'}^z c_{i'} R_{i'}^z, n_{j'}^z c_{j'} R_{j'}^z;\mu ;
          n_{i}^z c_{i} R_{i}^z, n_{j}^z c_{j} R_{j}^z)\ .
$$

The evident expression for integral
$$ j_{12}(n'a_1'R_1',m'a_2'R_2';\mu ;na_1R_1,ma_2R_2)= $$

\begin{equation}
\label{19}
=\int\limits_{-\infty}^{\infty}\int\limits_{-\infty}^{\infty}
   \psi_{n'}(x_1;a_1'R_1')\psi_{m'}(x_2;a_2'R_2')
   \exp\left\{-\frac{(x_1-x_2)^2}{\mu^2}\right\}
   \psi_{n}(x_1;a_1R_1)\psi_{m}(x_2;a_2R_2) dx_1dx_2
\end{equation}
is given in the Appendix (see eq.(\ref{A4})).

In the context of described above calculations it is suitable to
use the relation
\begin{equation}
\label{20} \frac{e^2}{r}=\frac{2e^2}{\sqrt{\pi}}\int\limits_0^1
    \exp\left( -\frac{r^2}{\mu^2_o}\right) \cdot
    \frac{d\tau}{(1-\tau^2)^{3/2}} \ ; \qquad
    \mu_o\equiv\frac{1}{\tau}\sqrt{1-\tau^2} \ .
\end{equation}
for evaluation of the Coulomb repulsion between protons. Thus,
\begin{equation}
\label{21} \frac{1}{<\Psi_{\nu}|\Psi_{\mu}>}\cdot
   <\Psi_{\nu}|\sum_{i>j=1}^Z\frac{e^2}{|\vec r_i-\vec r_j|}|\Psi_{\mu}>=
   \frac{2e^2}{\sqrt{\pi}} \sum_{i'>j'=1}^Z\sum_{i>j=1}^Z
   \int\limits_0^1 \frac{d\tau}{(1-\tau^2)^{3/2}}\times
\end{equation}

$$ \times<\psi_{\vec n_{i'}}(1)\psi_{\vec n_{j'}}(2)
   |\exp\left\{-\frac{|\vec r_1-\vec r_2|^2}{\mu_o^2}\right\}|
   \psi_{\vec n_i}(1)\psi_{\vec n_j}(2)>\times
$$

$$ \times\Biggl\{ 2
   \left( B^{-1}_{[\nu ,\mu ]}\right)_{ii'}
   \left( B^{-1}_{[\nu ,\mu ]}\right)_{jj'}-
   \left( B^{-1}_{[\nu ,\mu ]}\right)_{ij'}
   \left( B^{-1}_{[\nu ,\mu ]}\right)_{ji'}\Biggr\} \ ,
$$
i.e., in general, only integration over $\tau$ remains in the
equation (\ref{21}) for the Coulomb energy (all two-particle
integrals for Gaussian potential are expressed in terms of regular
functions).

The expression for center-of-mass kinetic energy has the form
\begin{equation}
\label{22} <\Psi_{\nu}|\hat T_{c.m.}|\Psi_{\mu}>=
   -\frac{\hbar^2}{2Am}\left(
   <\Psi_{\nu}|\sum_{i=1}^A\vec{\nabla}_i^2|\Psi_{\mu}>+
   <\Psi_{\nu}|\sum_{i\not= j=1}^A\vec{\nabla}_i\vec{\nabla}_j|
    \Psi_{\mu}> \right)  ,
\end{equation}
where the first term in the brackets is described by
eqs.(\ref{13}), (\ref{14}); the other can be evaluated using
general eq.(\ref{15}) for two-particle operators, i.e.
$$ \frac{1}{<\Psi_{\nu}|\Psi_{\mu}>}\cdot
   <\Psi_{\nu}|\sum_{i\not= j=1}^A\vec{\nabla}_i\vec{\nabla}_j|\Psi_{\mu}>=
$$

\begin{equation}
\label{23}
 =\sum_{i'\not=j'=1}^A\sum_{i\not=j=1}^A
   <\phi_{i'}^{[\nu ]}|\vec{\nabla}|\phi_i^{[\mu ]}>
   <\phi_{j'}^{[\nu ]}|\vec{\nabla}|\phi_j^{[\mu ]}>\times
\end{equation}
$$  \times \Biggl\{\left( B^{-1}_{[\nu ,\mu ]}\right)_{ii'}
   \left( B^{-1}_{[\nu ,\mu ]}\right)_{jj'}-
   \left( B^{-1}_{[\nu ,\mu ]}\right)_{ij'}
   \left( B^{-1}_{[\nu ,\mu ]}\right)_{ji'}\Biggr\} \ ;
$$
the explicit expressions for m.e. $<\phi_1|\vec{\nabla}|\phi_2>$
are given in the Appendix (see eq.(\ref{A3})).

Next, we briefly consider projection of the determinant function
on the state with defined angular momentum (the rotation effect in
phenomenological models). It is obvious that considered above
determinant functions $\Psi_{\nu}$ are superpositions of the
states $|JM>$. In order to obtain the state with defined angular
momentum $J$ and its projection $M$, we should define the
projector $\hat P^J_{MK}$ to operate on the total wave function
$\Psi_A$. It is suitable to write the projector $\hat P^J_{MK}$ in
the Hill-Wheeler's integral form

\begin{equation}
\label{24} \hat P_{MK}^J =\frac{2J+1}{8\pi^2}
  \int d\Omega D_{MK}^{J^{\ *}}(\Omega )\cdot\hat R(\Omega ) \ ,
\end{equation}
i.e. overlap integrals

\begin{equation}
\label{25} \frac{\int d\Omega D_{MM}^{J^{\ *}}(\Omega )
          <\Psi_A|\hat O|\hat R(\Omega )\Psi_A>}
     {\int d\Omega D_{MM}^{J^{\ *}}(\Omega )
          <\Psi_A|\hat R(\Omega )\Psi_A>} = \frac{
    \sum_{\nu\mu}w_{\nu}w_{\mu}\int d\Omega\cdot
     D_{MM}^{J^{\ *}}(\Omega )
          <\Psi_{\nu}|\hat O|\hat R(\Omega )\Psi_{\mu}>}
     {\sum_{\nu\mu}w_{\nu}w_{\mu}\int d\Omega D_{MM}^{J^{\ *}}(\Omega )
          <\Psi_{\nu}|\hat R(\Omega )\Psi_{\mu}>} \ ,
\end{equation}
should be evaluated to obtain eigenvalue of some physical operator
$\hat O$. Transformation of the spatial and spin-isospin functions
at rotation of the reference frames on Euler angles is performed
by usual way \cite{17}. A matrix $\tilde B$ can be defined
analogously to eq.(\ref{9}):

$$ \left( \tilde B_{[\nu ,\mu ]}\right)_{ij}\equiv
  \int\psi^{[\nu ]}_{\vec n_i}(\vec r;\vec a_i,\vec R_i)
  \hat R(\Omega )\psi^{[\mu ]}_{\vec n_j}(\vec r;\vec a_j,\vec R_j)
  d\vec r\sim
$$
\begin{equation}
\label{26} \sim\int\int\int dxdydz
  \left(\frac{x}{a_i}\right)^{n_i^x}
  \left(\frac{y}{b_i}\right)^{n_i^y}
  \left(\frac{z}{c_i}\right)^{n_i^z}
  \left(\frac{\tilde x}{a_j}\right)^{n_j^x}
  \left(\frac{\tilde y}{b_j}\right)^{n_j^y}
  \left(\frac{\tilde z}{c_j}\right)^{n_j^z}\times
\end{equation}
$$ \times\exp\left\{
  -\frac{(x-R_i^x)^2}{2a_i^2}
  -\frac{(y-R_i^y)^2}{2b_i^2}
  -\frac{(z-R_i^z)^2}{2c_i^2}
  -\frac{(\tilde x-R_j^x)^2}{2a_j^2}
  -\frac{(\tilde y-R_j^y)^2}{2b_j^2}
  -\frac{(\tilde z-R_j^z)^2}{2c_j^2} \right\} \ ,
$$
$\tilde x=\tilde x_1$, $\tilde y=\tilde x_2$, $\tilde z=\tilde
x_3$ are the Cartesian coordinates in the rotated frame of
reference that can be written using the rotation \cite{17} matrix
$||R_{m'm}(\Omega )||$, i.e.

\begin{equation}
\label{27}
 \tilde x_m=\sum^3_{m'=1} R_{m'm}(\Omega )\cdot x_{m'}\
,
  \qquad m=1,2,3.
\end{equation}
Integration in eq.(\ref{27}) can be performed completely by
transforming quadratic form from the exponent to diagonal one.

Note that two-particle matrix elements containing four different
orbitals can also be evaluated this way. Thus, in the case of m.e.

\begin{equation}
\label{28}
 < \psi_{\vec n_i}(\vec r_1;\vec a_i,\vec R_i)
  \psi_{\vec n_j}(\vec r_2;\vec a_j,\vec R_j)
   |\exp\left\{-\frac{(\vec r_1-\vec r_2)^2}{\mu^2}\right\}\hat R(\Omega ) |
  \psi_{\vec n_{i'}}(\vec r_1;\vec a_{i'},\vec R_{i'})
  \psi_{\vec n_{j'}}(\vec r_2;\vec a_{j'},\vec R_{j'}) >
\end{equation}
the quadratic form is constructed using 6-dimension vector $\vec
x=\vec x(x_1,x_2,\dots ,x_6)$ , where $\vec r_1=\vec
r_1(x_1,x_2,x_3)$, $\vec r_2=\vec r_2(x_4,x_5,x_6)$.

Hence, all integrations over particle coordinates can be performed
completely. Unfortunately, resulting expressions in the case of
the angular projection are too complicated to be written here.
Integration over Euler angles $\Omega
(\theta_1,\theta_2,\theta_3)$ have been carried out numerically,
although it is possible that further investigations will allow to
do it exactly.

\section{Numerical results.} \label{sect4}

There are few extremely significant systems in nuclear physics.
One of them is nucleus $^8$Be which have unique properties and is
 critical for the theory. Thus, $^8$Be is 8-nucleon
weak-bound system whose ground state is very narrow resonance near
the decay threshold. Therefore, calculation of the $^8$Be basic
spectroscopic characteristics will be serious test for the
considered microscopic model.

Due to the known symmetry of the nucleus $^8$Be we choose its wave
function in form

\begin{equation}
\label{29} \Psi_A(1,2,\dots ,8)=w_1\cdot\Psi_1+w_2\cdot
(\Psi_2+\Psi_3) \ ,
\end{equation}
where $\Psi_{\nu}$ is Slater determinant (\ref{6}) filled with
simple orbitals:

\noindent a) $\nu=1$
\begin{equation}
\label{30} \Psi_{\vec n_i}^{[\nu =1]}(j)=\left\{
   \begin{array}{ll}
   \psi_0(x_j;\tilde a_1,0)\psi_0(y_j;\tilde a_1,0)
   \psi_0(z_j;\tilde c_1,0), & \quad i=1,2,3,4; \\
   \psi_0(x_j;\tilde a_2,0)\psi_0(y_j;\tilde a_2,0)
   \psi_1(z_j;\tilde c_2,0), & \quad i=5,6,7,8; \\
   \end{array}  \right.
\end{equation}

\noindent b) $\nu=2,3$
\begin{equation}
\label{31} \Psi_{\vec n_i}^{[\nu =2]}(j)=\left\{
   \begin{array}{ll}
   \psi_0(x_j;a_1,0)\psi_0(y_j;a_1,0)\psi_0(z_j;c_1,-s\cdot R), &
                                        \ i=1,2,3,4; \\
   \psi_0(x_j;a_2,0)\psi_0(y_j;a_2,0)\psi_1(z_j;c_2,R), &
                                        \ i=5,6,7,8; \\
   \end{array}  \right.
\end{equation}

\begin{equation}
\label{32} \Psi_{\vec n_i}^{[\nu =3]}(j)=\left\{
   \begin{array}{ll}
   \psi_0(x_j;a_1,0)\psi_0(y_j;a_1,0)\psi_0(z_j;c_1,s\cdot R), &
                                        \ i=1,2,3,4; \\
   \psi_0(x_j;a_2,0)\psi_0(y_j;a_2,0)\psi_1(z_j;c_2,-R), &
                                        \ i=5,6,7,8; \\
   \end{array}  \right.
\end{equation}

The values $\{\tilde a_1,\tilde c_1,\tilde a_2,\tilde
c_2;a_1,c_1,a_2,c_2,R,s\}$ in these expressions are variation
parameters defined by minimization of the total nuclear energy
functional.

Let us consider our system at different values of the parameter
$R$ in order to examine the system dependence on the distance
between the fragments. The variation parameters were optimized for
every value of $R$; when $R\ge R_0$ the orbitals in the
determinant functions $\Psi_{\nu =2}$ and $\Psi_{\nu =3}$ are
continuously transformed in order to obtain wave functions of two
independent $\alpha$-particles at the limit $R\rightarrow\infty$,
i.e.

\begin{equation}
\label{33} \Psi_{\vec n_i}^{[\nu =2]}(j)=\left\{
   \begin{array}{ll}
   \psi_0(x_j;a_1,0)\psi_0(y_j;a_1,0)\psi_0(z_j;c_1,-s\cdot R), &
                                         i=1,2,3,4; \\ & \\
   -\alpha(\xi )\cdot\psi_0(x_j;a_1,0)\psi_0(y_j;a_1,0)
                 \psi_0(z_j;c_1,s\cdot R)+ & \\
   +\beta(\xi )\cdot\psi_0(x_j;a_2,0)\psi_0(y_j;a_2,0)
                 \psi_1(z_j;c_2,R), &
                                         i=5,6,7,8; \\
   \end{array}  \right.
\end{equation}

\begin{equation}
\label{34} \Psi_{\vec n_i}^{[\nu =3]}(j)=\left\{
   \begin{array}{ll}
   \psi_0(x_j;a_1,0)\psi_0(y_j;a_1,0)\psi_0(z_j;c_1,s\cdot R), &
                                         i=1,2,3,4; \\ & \\
   \alpha(\xi )\cdot\psi_0(x_j;a_1,0)\psi_0(y_j;a_1,0)
                 \psi_0(z_j;c_1,-s\cdot R)+ & \\
   +\beta(\xi )\cdot\psi_0(x_j;a_2,0)\psi_0(y_j;a_2,0)
                 \psi_1(z_j;c_2,-R), &
                                         i=5,6,7,8; \\
   \end{array}  \right.
\end{equation}

The functions $\alpha (\xi )$, $\beta (\xi )$ are defined by

\begin{equation}
\label{35} \alpha(\xi )=\frac{\xi}{\sqrt{1+2\xi\sqrt{1-\xi^2}}} \
,\qquad \beta(\xi
)=\frac{\sqrt{1-\xi^2}}{\sqrt{1+2\xi\sqrt{1-\xi^2}}} \ ,\qquad
\xi\equiv\frac{R-R_0}{\eta_0\cdot R_0} \ , \qquad 0\le\xi\le 1.
\end{equation}
Choice of $R_0$ and $\eta_0$ in eq.(\ref{36}) accords to choice of
interval $[R_0,(\eta_0+1)R_0]$ where one of the $\alpha$-clusters
"loses" its excitation. Next we define the antisymmetrization of
the wave function, if $R\le R_0$ then the Pauli principle acts on
all A nucleons in the nucleus. In the range $[R_0,(\eta_0+1)R_0]$
the antisymmetrization between particles in the different clusters
is gradually "switched off". Moreover, both "losing" of the
excitation and "switching off" of the Pauli principle action are
simulated by the same functions $\alpha (\xi )$, $\beta (\xi )$.

The total wave function of the system in the considered range
$[R_0,(\eta_0+1)R_0]$ is presented by

\begin{equation}
\label{36} \Psi_A(1,2,\dots ,8)\sim\beta(\xi )\cdot w_1^0\Psi_1+
   [1+\alpha(\xi )]\cdot w_2^0\cdot (\Psi_2+\Psi_2) \ ,
\end{equation}
where $w_1^0$, $w_2^0$ are weights of the according determinant
functions at $R=R_0$. The parameter $\alpha_S$ in the microscopic
Hamiltonian (see eq.(\ref{4})) can be defined by

\begin{equation}
\label{37} \alpha_S=\alpha_S(R)=\left\{
   \begin{array}{ll}
   0,            & \quad R<R_0 \ , \\
   \alpha(\xi ), & \quad R_0\le R\le (\eta_0+1)R_0 \ , \\
   1,            & \quad R>(\eta_0+1)R_0 \ . \\
   \end{array}  \right.
\end{equation}

In order to evaluate the overlap integrals corresponding to some
physical quantity it is necessary to calculate the matrix $B_{[\nu
,\mu ]}$ and the inverse matrix $B^{-1}_{[\nu ,\mu ]}$. In the
case of nucleus $^8$Be these matrices are in general

\begin{equation}
\label{38} B_{[\nu ,\mu]}=
   \begin{array}{|cccccccc|}
   \alpha_{\nu\mu} &0&0&0& \gamma_{\nu\mu} &0&0&0  \\
   0& \alpha_{\nu\mu} &0&0&0& \gamma_{\nu\mu} &0&0 \\
   0&0& \alpha_{\nu\mu} &0&0&0& \gamma_{\nu\mu} &0 \\
   0&0&0& \alpha_{\nu\mu} &0&0&0& \gamma_{\nu\mu}  \\
   \delta_{\nu\mu} &0&0&0& \beta_{\nu\mu} &0&0&0  \\
   0& \delta_{\nu\mu} &0&0&0& \beta_{\nu\mu} &0&0 \\
   0&0& \delta_{\nu\mu} &0&0&0& \beta_{\nu\mu} &0 \\
   0&0&0& \delta_{\nu\mu} &0&0&0& \beta_{\nu\mu}  \\
   \end{array} \ ;
\qquad
 \det\Biggl{\|}\left( B_{[\nu ,\mu ]}\right)_{ij}\Biggr{\|}=
  \zeta_0^4 \ ;
\end{equation}

$$ \zeta_0\equiv
   \alpha_{\nu\mu}\beta_{\nu\mu}-\gamma_{\nu\mu}\delta_{\nu\mu} \ ;
$$

\begin{equation}
\label{39} B_{[\nu ,\mu]}^{-1}=
   \begin{array}{|cccccccc|}
   \beta_{\nu\mu}/\zeta_0 &0&0&0& -\gamma_{\nu\mu}/\zeta_0 &0&0&0  \\
   0& \beta_{\nu\mu}/\zeta_0 &0&0&0& -\gamma_{\nu\mu}/\zeta_0 &0&0 \\
   0&0& \beta_{\nu\mu}/\zeta_0 &0&0&0& -\gamma_{\nu\mu}/\zeta_0 &0 \\
   0&0&0& \beta_{\nu\mu}/\zeta_0 &0&0&0& -\gamma_{\nu\mu}/\zeta_0  \\
   -\delta_{\nu\mu}/\zeta_0 &0&0&0& \alpha_{\nu\mu}/\zeta_0 &0&0&0  \\
   0& -\delta_{\nu\mu}/\zeta_0 &0&0&0& \alpha_{\nu\mu}/\zeta_0 &0&0 \\
   0&0& -\delta_{\nu\mu}/\zeta_0 &0&0&0& \alpha_{\nu\mu}/\zeta_0 &0 \\
   0&0&0& -\delta_{\nu\mu}/\zeta_0 &0&0&0& \alpha_{\nu\mu}/\zeta_0  \\
   \end{array} \ .
\end{equation}
Evident expressions for $\alpha_{\nu\mu}$, $\beta_{\nu\mu}$,
$\gamma_{\nu\mu}$, $\delta_{\nu\mu}$ can be obtained by using the
definitions (\ref{30})--(\ref{34}) (see eqs.(\ref{A6}) in the
Appendix).

Concrete calculations on the basis of the above considered
equations have been performed for the case of some well known
variants of the central exchange NN-potentials. Thus, the results
for Volkov's NN-potential \cite{18} (the 1st variant case) are
accurate. Some of the results are given in Table \ref{tab1}. There
are energy $E_{0^+}$ of the $^8$Be ground $0^+$-state, energy
$U_{Coul}$ of the Coulomb repulsion between protons, r.m.s. matter
radii $<r^2>^{1/2}$ as well as optimized values of the variation
parameters $\{a_i,b_i,c_i,\tilde a_i,\tilde b_i,\tilde c_i,R,s\}$
in the determinant functions $\Psi_{\nu}$ and its non-normalized
weights $\{w_i\}$. In the case of N1 the wave function (\ref{30})
contains 10 nonlinear variation parameters and describes both
orbitals "polarization" and cluster structure of the nucleus
$^8$Be. In the case of N2, the superposition of two determinant
functions

\begin{equation}
\label{40} \Psi_A=(1,2,\dots ,A)=w_1\cdot\Psi_1(\{\vec r_i\};
   \tilde a_1=\tilde b_1,\tilde c_1;
   \tilde a_2=\tilde b_2,\tilde c_2)+
\end{equation}
$$
  +w_4\cdot\Psi_4(\{\vec r_i\};
   \tilde{\tilde a}_1=\tilde{\tilde b}_1,\tilde{\tilde c}_1;
   \tilde{\tilde a}_2=\tilde{\tilde b}_2,\tilde{\tilde c}_2).
$$
taking into account only orbitals "polarization" has been used.
Column N3 corresponds to the previous case with neglecting of the
deformation, i.e.

\begin{equation}
\label{41} \Psi_A=(1,2,\dots ,A)=w_1\cdot\Psi_1(\{\vec
r_i\};\tilde a)+
                      w_4\cdot\Psi_4(\{\vec r_i\};\tilde{\tilde a}).
\end{equation}
The values of column N4 are calculated on the basis of one
determinant function with different degree of "polarization" for
s- and p-orbitals (4 variation parameters: $\tilde a_1=\tilde
b_1$, $\tilde c_1$, $\tilde a_2=\tilde b_2$, $\tilde c_2$). In the
case of N5 we have used one determinant function with the same
deformation for s- and p-orbitals (2 parameters: $\tilde a=\tilde
b$, $\tilde c$). The simplest approach is microscopic SU(3)-model
with one variation parameter $\tilde a=\tilde b=\tilde c$ that
corresponds to column N6. The data in column N7 have been
calculated using the most complicated wave function $\Psi_A$. It
differs from one of the case N1 by additional shell-type
determinant function $\Psi_{\nu =4}$ filled by one-particle states
containing independent variation parameters $\tilde{\tilde
a}_1=\tilde{\tilde b}_1$, $\tilde{\tilde c}_1$, $\tilde{\tilde
a}_2=\tilde{\tilde b}_2$, $\tilde{\tilde c}_2$. This approach
employs 14 nonlinear variation parameters. And, finally, data in
the column N8 have been evaluated using the previous function
$\Psi_A$, but without angular projection.

First we consider numerical results obtained with one-center
functions $\Psi_A$ (models N2--N6). The data in Table \ref{tab1}
show that the most binding energy growth is caused by taking into
account the oscillator field deformation (compare models N6 and
N5). Independent orbital "polarization" is also significant
(compare models N5 and N4). If the nonlinear variation parameters
$\{ a_i,b_i,c_i\}$ are the same (i.e. deformation is neglected)
than expanding of the basis from 1 to 2 determinant functions
$\Psi_{\nu}$ does not practically change the nuclear energy E
(comp. models N3 and N6). Moreover, our estimations show that the
employment of two, three and more determinant functions
$\Psi_{\nu}$ with the same SU(3)-symmetry does not allow to reach
significant advance even if all considered s- and p-orbital
deformations are taken into account (comp. models N2 and N4).
Hence, the deformation effects can be considered using only one
one-center Slater determinant. More effective is expansion of the
variation basis by cluster wave function. It enhances the $^8$Be
nuclear energy by approximately 3 MeV (comp. models N1, N4, N2).

Fig. \ref{fig1} shows $^8$Be mass distribution $\rho(\vec r)$
evaluated by using the wave function (\ref{30}). The optimum
variation parameters correspond to model N1 (see Table
\ref{tab1}). The calculations are carried out in the nuclear
center-of-mass reference frames. Surface $(x,y=0,z)$ is displayed
in Fig. \ref{fig1} due to axial symmetry of the distribution
$\rho(\vec r)$. We can see from Fig. \ref{fig1} that distance
between density maximums (centers of mass for two
$\alpha$-particles) is $D\approx 4.2$ fm. Density of junction
between $\alpha$-particles is approximately 0.7$\rho_{max}$. The
distribution $\rho(\vec r)$ obtained in the framework of the other
considered models is rather similar to Fig. \ref{fig1}, although
distance $D$ and width of junction can be essentially different.

Evaluated parameter $R$ dependence of the $^8$Be total energy
functional $f_E(R,\{a_i,b_i,c_i\})$ is plotted in Fig. \ref{fig2}.
The shown dependence have been calculated under condition that for
every value $R$ functional $f_E$ is optimized to the other
variation parameters. Such computations can be useful for
estimations of the nucleus-nucleus potentials. Note that in a
range where clusters overlap each other, the distance $D$ between
density maximums can essentially differ from its asymptotic value
which is equal to $D_{as}=R(1+s)$ in our case. Fig. \ref{fig3}
shows dependence $D=f(R)$ obtained for model N1. We can see that
the distance $D$ can slightly decrease at $0\le R\le 1.5$ fm and
only at $R>2$ fm the $R$ dependence of $D$ is almost linear.

Table \ref{tab2} contains characteristics of the $^6$He ground
$0^+$-state calculated using three variants of NN-potentials: V1
from Ref. \cite{18}, BB1 from Ref. \cite{19} and G5 from Ref.
\cite{20}. Computations are carried out in the framework of two
models. Wave function of model I is a Slater determinant for 6
nucleons filled by orbitals (\ref{31}), and contains four
variation parameters: $\tilde a_1=\tilde b_1$, $\tilde c_1$ for
first four orbitals and $\tilde a_2=\tilde b_2$, $\tilde c_2$ for
the last two orbitals. Wave function of the second model is
superposition of a few Slater determinants, similarly to the
$^8$Be: one-center determinant (model~I) and two-center
determinants with centers in points $\vec R_1=(0,0,\pm R)$ and
$\vec R_2=(0,0,\mp sR)$. This model contains 10 nonlinear
variation parameters whose optimum values for the considered
variants of NN-forces are given in Table \ref{tab2}.

It should be remembered, that a core in $^6$He is nucleus $^4$He.
Its energy calculated using simple SU(3)-model function is equal
for three variants of NN-forces: $E_{\alpha}(\mbox{V1})=-27.09$
MeV, $E_{\alpha}(\mbox{BB1})=-27.37$ MeV,
$E_{\alpha}(\mbox{G5})=-28.29$ MeV. Comparison of these values
with $^6$He ground state energies in Table \ref{tab2} shows that
two neutrons are bound only in the case of model II and
NN-potentials V1 and G5. Therefore, we can conclude that wave
function of model II describes the real systems more accurate than
one-center function of model I.

Note that authors don't claim superexact description of the
considered here nuclear systems which has been already studied
profoundly by many authors (see, for example, Refs. \cite{9,10}).
The purpose of this study is to clarify the question about
significance of "compact" and "loosely bound" structures at
consideration of nuclear states using simple trial wave
functions.

Performed analysis allows to draw up some general scheme for the
construction of the simplest and quite flexible trial wave
function of weakly bound state as well as a strategy for
concrete computation. First, possible significance of the
deformation in formation of considered nucleus state is roughly
estimated in the framework of the one-center determinant function
model. Next, cluster degrees of freedom are taken into account.
Hierarchy of the nucleon motion modes can be determined using
non-projected wave functions $\Psi_A$ that significantly reduces
computations. Moreover, such reduced evaluations give approximate
optimum values for almost all variation parameters (for example,
compare corresponding parameters of models N7 and N8 in Table
\ref{tab1}). It is obvious that this fact is especially
significant in order to find the global minimum of the total
nuclear energy functional by employing projected functions
$\Psi_A^{JM}(1,2,\dots ,A)$.

\appendix{}
\section*{}

The evident expression (\ref{7}) allows to calculate:

\noindent 1) partial m.e. (\ref{10})

\begin{equation}
\label{A1} g_{n_1n_2}(a_1,R_1;a_2,R_2)=
   \frac{g_0}{N\left( n_1;\frac{R_1}{a_1}\right)
              N\left( n_2;\frac{R_2}{a_2}\right)}\cdot
   \left(\frac{Z_0}{X_0Y_0}\right)^{n_1+n_2}\times
\end{equation}

$$ \times F\left( -\frac{n_1+n_2}{2},
         -\frac{n_1+n_2}{2}+\frac{1}{2};0;\frac{1}{Z_0^2}\right) \ ,
$$

$$ g_0\equiv\sqrt{X_0Y_0}\cdot\exp\left\{
   -\frac{(R_1-R_2)^2}{2(a_1^2+a_2^2)}\right\} \ ;\qquad
X_0=\sqrt{\frac{2a_2^2}{a_1^2+a_2^2}}\ ;\qquad
Y_0=\sqrt{\frac{2a_1^2}{a_1^2+a_2^2}}\ ;
$$

$$
Z_0=\frac{1}{2}\left(\frac{R_1}{a_1} X_0+
    \frac{R_2}{a_2} Y_0\right) \ ;
$$

\noindent 2) m.e. (\ref{14}) used for evaluation of the kinetic
energy

$$ t_{n'n}(a',R';a,R)\equiv <\psi_{n'}(x;a',R')|
   -\frac{\hbar^2}{2m}\frac{\partial^2}{\partial x^2}|
   \psi_{n}(x;a,R)>=
$$

\begin{equation}
\label{A2} =-\frac{\hbar^2}{2ma^2}\cdot\frac{1}{N\left(
n;\frac{R}{a}\right)}
   \Biggl\{\Biggr. n(n-1)N\left( n-2,\frac{R}{a}\right)
   g_{n',n-2}(a',R';a,R)+
\end{equation}

$$ +2n\frac{R}{a}N\left( n-1,\frac{R}{a}\right)
   g_{n',n-1}(a',R';a,R)
+\left(\frac{R^2}{a^2}-2n-1\right)N\left( n,\frac{R}{a}\right)
   g_{n',n}(a',R';a,R)-
$$

$$-2\frac{R}{a}N\left( n+1,\frac{R}{a}\right)
   g_{n',n+1}(a',R';a,R)+
   N\left( n+2,\frac{R}{a}\right) g_{n',n+2}(a',R';a,R)
   \Biggl.\Biggr\} ;
$$

\vspace{3mm}
\begin{equation}
\label{A3} <\phi_1|\vec{\nabla}|\phi_2>=
   <\phi_1|\frac{\partial}{\partial x}\phi_2>\cdot\vec e_x+
   <\phi_1|\frac{\partial}{\partial y}\phi_2>\cdot\vec e_y+
   <\phi_1|\frac{\partial}{\partial z}\phi_2>\cdot\vec e_z \ ,
\end{equation}

\vspace{3mm}
\begin{equation}
\label{A4}
<\phi_1|\frac{\partial}{\partial x}\phi_2>\equiv
   <\psi_{\vec{\tilde n}}(\vec r;\vec{\tilde a},\vec{\tilde R})|
    \frac{\partial}{\partial x}\psi_{\vec n}(\vec r;\vec a,\vec R)>
=g_{\tilde n_yn_y}(\tilde b,\tilde R^y;bR^y)
   g_{\tilde n_zn_z}(\tilde c,\tilde R^z;cR^z)\times
\end{equation}

$$ \times\Biggl[ n^x\frac{N(n^x;R^x/a)}{N(n^x-1;R^x/a)}
   g_{\tilde n^x,n^x-1}(\tilde a,\tilde R^x;aR^x)+
   \frac{R^x}{a}g_{\tilde n^x,n^x}(\tilde a,\tilde R^x;aR^x)-
$$

$$
-\frac{N(n^x;R^x/a)}{N(n^x+1;R^x/a)}
   g_{\tilde n^x,n^x+1}(\tilde a,\tilde R^x;aR^x)\Biggr] ,
$$
equations for $<\phi_1|\frac{\partial}{\partial y}\phi_2>$ and
$<\phi_1|\frac{\partial}{\partial z}\phi_2>$ can be obtained from
(\ref{A3}) after obvious substitution $a\rightarrow b$,
$R^x\rightarrow R^y$ and $a\rightarrow c$, $R^x\rightarrow R^z$;

\noindent 3) 2-dimension integral (\ref{19}) determining the
potential energy of NN-interaction

\vspace{3mm}
\begin{equation}
\label{A5}
j_{12}(\tilde n\tilde a_1\tilde R_1,\tilde m\tilde a_2\tilde
R_2;
  \mu ;n a_1 R_1, m a_2 R_2)=
  \frac{1}{\sqrt{\tilde a_1\tilde a_2a_1a_2}}\cdot
  \frac{\mu^2}{\eta_0}\times
\end{equation}

$$ \times\left(\frac{1}{2\tilde a_1}\right)^{\tilde n}
   N\left( \tilde n;\frac{\tilde R_1}{\tilde a_1}\right)
   \left(\frac{1}{2\tilde a_2}\right)^{\tilde m}
   N\left( \tilde m;\frac{\tilde R_2}{\tilde a_2}\right)
   \left(\frac{1}{2 a_1}\right)^{ n}
   N\left( n;\frac{ R_1}{ a_1}\right)
   \left(\frac{1}{2 a_2}\right)^{ m}
   N\left( m;\frac{ R_2}{ a_2}\right)\times
$$

$$ \times\left(\frac{\mu}{\xi_0\eta_0}\right)^{\tilde n+n}
   \left(\eta\frac{\xi_0}{\eta_0}\right)^{\tilde m+m}
   \cdot\exp\{+arg\}
\sum_{l-0}^{\tilde n+n}
   \frac{\eta_0^l(\tilde n+n)!}{l!(\tilde n+n-l)!}
   \sum_{k=0}^{\left[\frac{l}{2}\right]}
   \frac{l!}{k!(l-2k)!}\times
$$

$$
   \times\Biggl[\frac{\mu}{\xi_0}\left(\frac{R_1}{a_1^2}+
     \frac{\tilde R_1}{\tilde a_1^2}\right)\Biggr]^{l-2k}
\sum_{i=0}^{\left[\frac{\tilde n+n+
                      \tilde m+m-l}{2}\right]}
   \frac{(\tilde n+n+\tilde m+m-l)!}
        {i!(\tilde n+n+\tilde m+m-l-2i)!}\times
$$

$$ \times
\Biggl[\frac{\mu}{\xi_0\eta_0}\left(\frac{R_1}{a_1^2}+
     \frac{\tilde R_1}{\tilde a_1^2}\right)+
     \frac{\mu\xi_0}{\eta_0}\left(\frac{R_2}{a_2^2}+
     \frac{\tilde R_2}{\tilde a_2^2}\right)\Biggr]
     ^{\tilde n+n+\tilde m+m-l-2i}\ ;
$$
where

\begin{equation}
\label{A6} \xi_0=\left[
1+\frac{\mu^2}{2a_1^2}+\frac{\mu^2}{2a_2^2}
      \right]^{\frac{1}{2}}\ ,
\end{equation}

$$
\eta_0=\left[\frac{\mu^2}{2a_1^2}+\frac{\mu^2}{2\tilde a_1^2}+
             \frac{\mu^2}{2a_2^2}+\frac{\mu^2}{2\tilde a_2^2}+
      \Biggl(\frac{\mu^2}{2a_1^2}+\frac{\mu^2}{2\tilde a_1^2}\Biggr)
      \Biggl(\frac{\mu^2}{2a_2^2}+\frac{\mu^2}{2\tilde a_2^2}\Biggr)
       \right]^{\frac{1}{2}} \ ,
$$

$$ arg=-\frac{1}{2}\left(\frac{\tilde R_1^2}{\tilde a_1^2}+
                      \frac{R_1^2}{a_1^2}+
                      \frac{\tilde R_2^2}{\tilde a_2^2}+
                      \frac{R_2^2}{a_2^2}\right) +
$$

$$ +\frac{1}{4}\frac{
   \left( 1+\frac{\mu^2}{2\tilde a_2^2}+\frac{\mu^2}{2a_2^2}\right)
   \left(\frac{\mu\tilde R_1}{\tilde a_1^2}+
         \frac{\mu R_1}{a_1^2}\right)^2+
   \left( 1+\frac{\mu^2}{2\tilde a_1^2}+\frac{\mu^2}{2a_1^2}\right)
   \left(\frac{\mu\tilde R_2}{\tilde a_2^2}+
         \frac{\mu R_2}{a_2^2}\right)^2  }{
   \left( 1+\frac{\mu^2}{2\tilde a_1^2}+\frac{\mu^2}{2a_1^2}\right)
   \left( 1+\frac{\mu^2}{2\tilde a_2^2}+\frac{\mu^2}{2a_2^2}\right)
   -1} \ +
$$

$$ +\frac{1}{2}\frac{
    \left(\frac{\mu\tilde R_1}{\tilde a_1^2}+
         \frac{\mu R_1}{a_1^2}\right)
    \left(\frac{\mu\tilde R_2}{\tilde a_2^2}+
         \frac{\mu R_2}{a_2^2}\right) }{
   \left( 1+\frac{\mu^2}{2\tilde a_1^2}+\frac{\mu^2}{2a_1^2}\right)
   \left( 1+\frac{\mu^2}{2\tilde a_2^2}+\frac{\mu^2}{2a_2^2}\right)
   -1} \ .
$$

Due to definitions (\ref{28})--(\ref{32}) the $B_{[\nu ,\mu ]}$
matrix elements (see eq.(\ref{36})) can be expressed in terms of
functions $g_{nm}$ from eq.(\ref{A1}):

\begin{equation}
\label{A7}
\begin{array}{cc}
\alpha_{11}=\alpha_{22}=1, &
\alpha_{12}=\alpha_{21}=g_{00}^2(\tilde a_1,0;\tilde a_{11},0)
                        g_{00}(\tilde c_1,0;\tilde c_{11},0), \\
\beta_{11}=\beta_{22}=1, & \beta_{12}= \beta_{21}= g_{00}^2(\tilde
a_2,0;\tilde a_{22},0)
                        g_{11}(\tilde c_2,0;\tilde c_{22},0), \\
\gamma_{11}=\gamma_{22}=0, &
\gamma_{12}=\gamma_{21}=g_{00}^2(\tilde a_1,0;\tilde a_{22},0)
                        g_{01}(\tilde c_1,0;\tilde c_{22},0), \\
\delta_{11}=\delta_{22}=0, &
\delta_{12}=\delta_{21}=g_{00}^2(\tilde a_2,0;\tilde a_{11},0)
                        g_{10}(\tilde c_2,0;\tilde c_{11},0). \\
\end{array}
\end{equation}

\begin{figure}[tbp]
\caption{Mass density of $^8$Be as a function of the variables $z$
and $r_{\bot}=\sqrt{x^2+y^2}$, and contours of constant mass
density in the section of $^8$Be by the ($x,z$) plane in the
intrinsic coordinate frame rigidly bound to the nucleus.}
\label{fig1}
\end{figure}

\begin{figure}[tbp]
\caption{Total energy of the $^8$Be nucleus as a function of the
variational parameter $R$.}
\label{fig2}
\end{figure}

\begin{figure}[tbp]
\caption{$2\alpha$-distance as a function of the parameter $R$.}
\label{fig3}
\end{figure}

\begin{table}
\caption{The nucleus $^8$Be ground state in microscopic models,
Volkov's NN-potential \protect\cite{16}. The models (N1--N8) are
described in the text.} \label{tab1}
\begin{tabular}{cccccccccc}
\multicolumn{2}{c}{Characteristics}
     & N 1 & N 2 & N 3 & N 4 & N 5 & N 6 & N 7 & N 8 \\
\hline \multicolumn{2}{c}{$E$, MeV}
     & -52.732 & -50.017 & -36.859 & -49.605 & -47.910 & -36.135 & -52.975 & -44.836 \\
\multicolumn{2}{c}{$U_{Coul}$, MeV}
     & 3.145   & 3.339   & 3.422   & 3.353   & 3.366   & 3.446   & 3.154   & 3.279   \\
\multicolumn{2}{c}{$<r^2>^{1/2}$, fm}
                          & 2.52 & 2.33 & 2.19 & 2.31 & 2.30 & 2.15 & 2.52 & 2.38 \\
& $\tilde a_1=\tilde b_1$ & 1.27 & 1.22 & 1.49 & 1.29 & 1.32 &
1.60 & 1.22 & 1.27 \\
%\cline{2-10}
$\Psi_1$ & $\tilde c_1$   & 2.23 & 2.33 & 1.49 & 2.43 & 2.07 &
1.60 & 2.18 & 1.89 \\
%\cline{2-10}
& $\tilde a_2=\tilde b_2$ & 1.35 & 1.29 & 1.49 & 1.35 & 1.32 &
1.60 & 1.29 & 1.34 \\
%\cline{2-10}
         & $\tilde c_2$   & 1.91 & 1.80 & 1.49 & 1.93 & 2.07 & 1.60 & 1.79 & 1.62 \\
& $a_1=b_1$               & 1.49 &      &      &      &      &
& 1.47 & 1.57 \\
%\cline{2-10}
& $c_1$                   & 1.59 &      &      &      &      &
& 1.56 & 1.56 \\
%\cline{2-10}
$\Psi_2,$ & $a_2=b_2$     & 1.28 &      &      &      &      &
& 1.25 & 1.30 \\
%\cline{2-10}
$\Psi_3$  & $c_2$         & 1.37 &      &      &      &      &
& 1.36 & 1.36 \\
%\cline{2-10}
& $s$                     & 1.28 &      &      &      &      &
& 1.29 & 1.30 \\
%\cline{2-10}
& $R$, fm                 & 1.76 &      &      &      &      &
& 1.73 & 1.53 \\ & $\tilde {\tilde a}_1=\tilde {\tilde b}_1$
                          &      & 1.50 & 1.86 &      &      &      & 1.50 & 1.51 \\
%\cline{2-10}
$\Psi_4$ & $\tilde {\tilde c}_1$
                          &      & 2.70 & 1.86 &      &      &      & 2.48 & 2.10 \\
%\cline{2-10}
& $\tilde {\tilde a}_2=\tilde {\tilde b}_2$
                          &      & 1.53 & 1.86 &      &      &      & 1.65 & 1.72 \\
%\cline{2-10}
         & $\tilde {\tilde c}_2$
                          &      & 2.28 & 1.86 &      &      &      & 2.37 & 2.12 \\
\multicolumn{2}{c}{$w_1$}&1.74& 1.35 & 1.03 & 1.00 & 1.00 & 1.00 &
0.97 & 0.50 \\ \multicolumn{2}{c}{$w_2=w_3$}&1.00&  &      &
&      &      & 0.69 & 0.21 \\ \multicolumn{2}{c}{$w_4$}&    &
0.72 & 0.69 &      &      &      & 0.42 & 0.34 \\
\end{tabular}
\end{table}

\begin{table}
\caption{Calculated $^6$He features. Energies $E$, $U_{Coul}$ are
given in MeV, rms radii, $\{a_i,b_i,c_i\}$, $R$ are given in fm.}
\label{tab2}
\begin{tabular}{cccccccc}
\multicolumn{2}{c}{} & \multicolumn{2}{c}{Poten. V1
\protect\cite{16}} & \multicolumn{2}{c}{Poten. BB1
\protect\cite{18}} & \multicolumn{2}{c}{Poten. G5
\protect\cite{19}} \\ \multicolumn{2}{c}{Characteristics} &
  I model & II model & I model & II model & I model & II model \\
\hline \multicolumn{2}{c}{$E$}
     & -25.296 & -27.396 & -24.451 & -26.918 & -29.636 & -30.831 \\
\multicolumn{2}{c}{$U_{Coul}$}
     & 0.7139  & 0.7633  & 0.6820  & 0.7428  & 0.7300  & 0.7591  \\
\multicolumn{2}{c}{$r_{rms}^{matter}$}
                          & 2.51 &      & 2.64 &      & 2.47 &  \\
\multicolumn{2}{c}{$r_{rms}^{proton}$}
                          & 1.86 &      & 1.95 &      & 1.78 &  \\
\multicolumn{2}{c}{$r_{rms}^{neutron}$}
                          & 2.78 &      & 2.92 &      & 2.76 &  \\
& $\tilde a_1=\tilde b_1$ & 1.31 & 1.28 & 1.36 & 1.34 & 1.34 &
1.33 \\
%\cline{2-8}
$\Psi_{\nu =1}$
           & $\tilde c_1$ & 2.19 & 2.05 & 2.32 & 2.16 & 1.97 & 1.85 \\
%\cline{2-8}
& $\tilde a_2=\tilde b_2$ & 1.45 & 1.36 & 1.55 & 1.47 & 1.60 &
1.74 \\
%\cline{2-8}
         & $\tilde c_2$   & 2.14 & 2.05 & 2.23 & 2.16 & 2.00 & 1.83 \\
& $a_1=b_1$               &      & 1.39 &      & 1.41 &      &
1.39 \\
%\cline{2-8}
& $c_1$                   &      & 1.37 &      & 1.39 &      &
1.41 \\
%\cline{2-8}
$\Psi_{\nu =2,3}$ & $a_2=b_2$ &  & 1.86 &      & 1.85 &      &
1.46 \\
%\cline{2-8}
& $c_2$                   &      & 2.54 &      & 2.56 &      &
2.47 \\
%\cline{2-8}
& $s$                     &      & 1.16 &      & 1.13 &      &
1.65 \\
%\cline{2-8}
& $R$                     &      & 1.31 &      & 1.44 &      &
0.91 \\ \multicolumn{2}{c}{$w_1$} & 1 & 0.9209 & 1 & 0.8509 & 1 &
0.8993 \\ \multicolumn{2}{c}{$w_2$} & 0 & 0.6223 & 0 & 0.7033 & 0
& 0.4835 \\
\end{tabular}
\end{table}

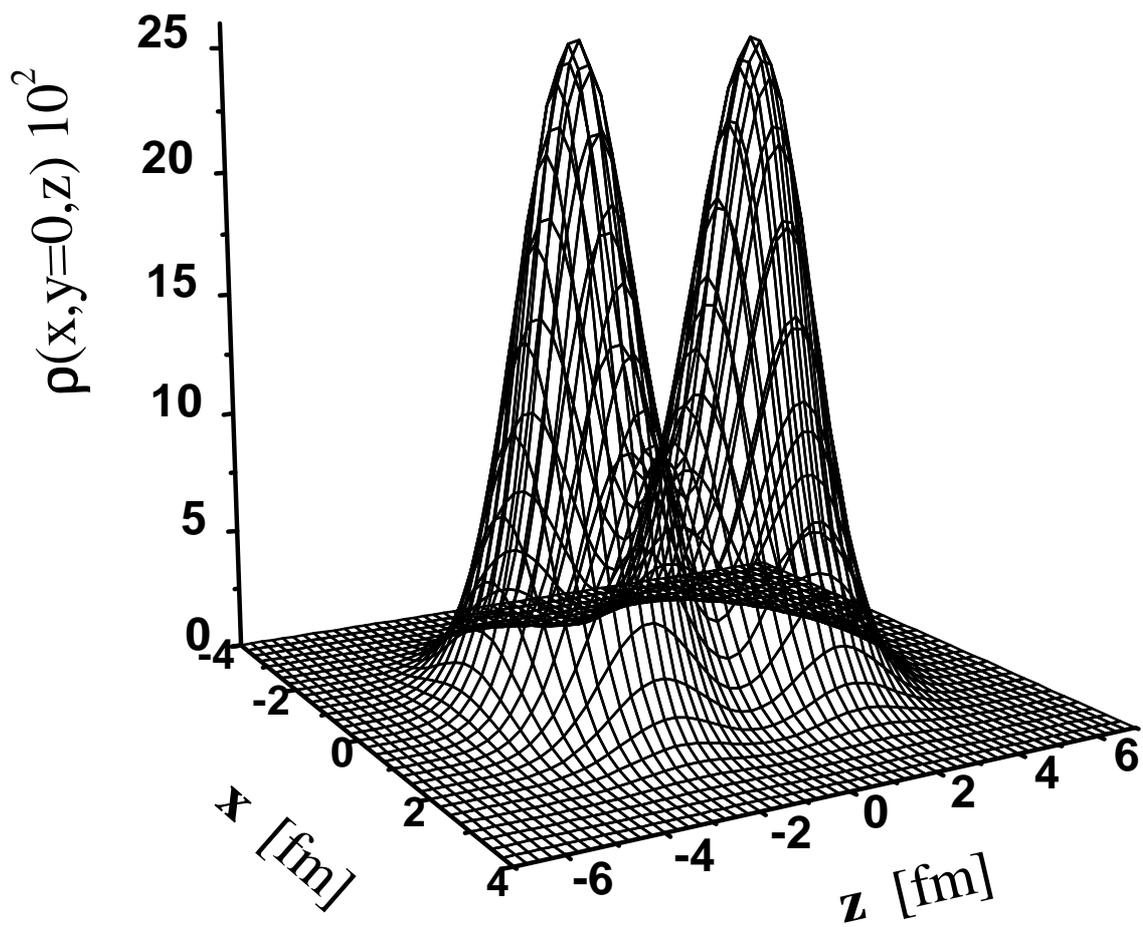
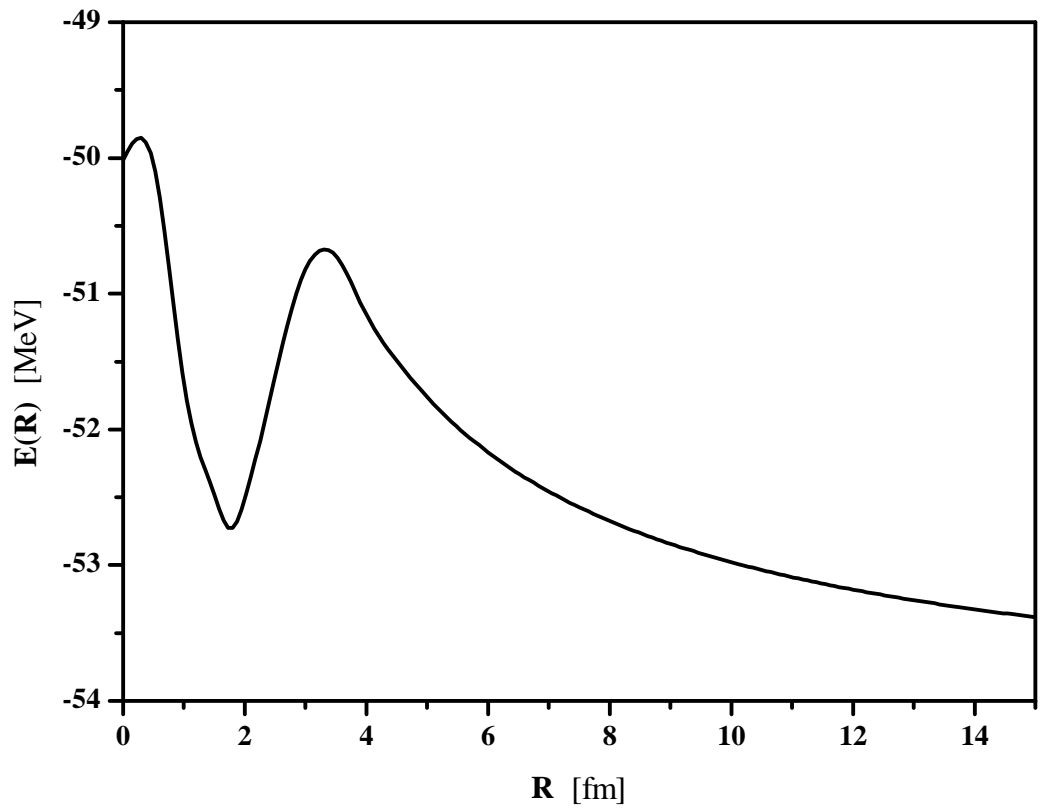
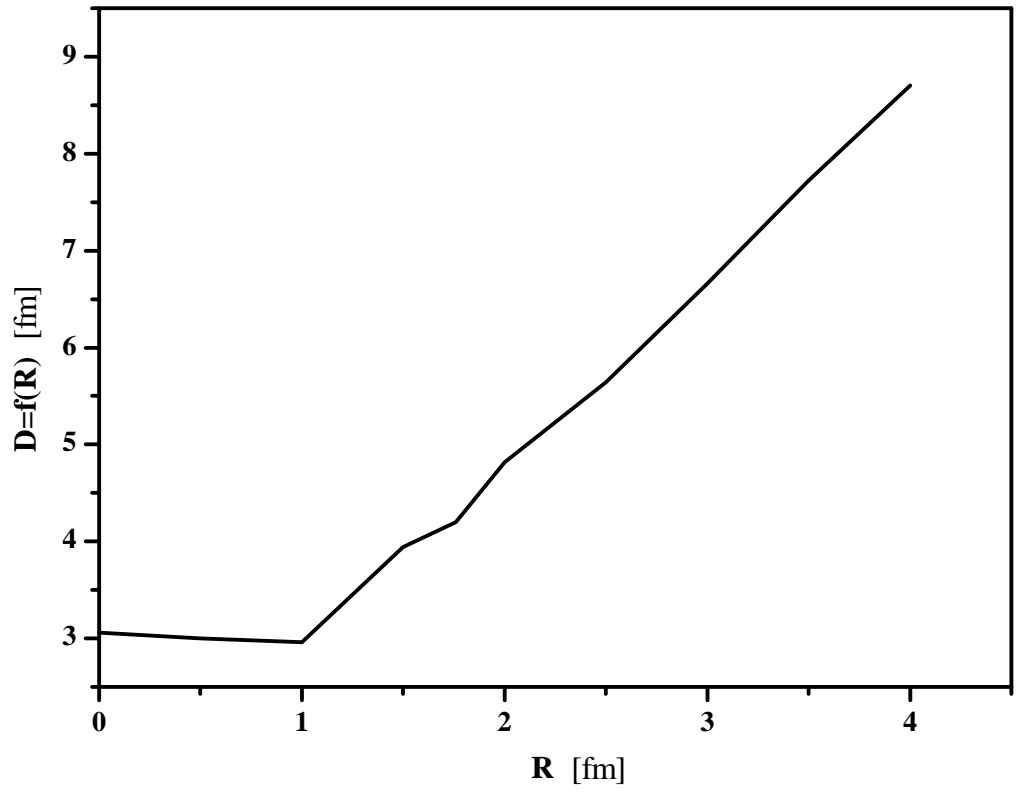
\end{document}